\documentclass{an}
\usepackage{graphicx}
\usepackage{times}
\usepackage{fancyhdr}
\begin{document}

\title{The origin of magnetic fields in galaxies:\\
observational tests with the Square Kilometre Array}

\author{Rainer~Beck}

\institute{Max-Planck-Institut f\"ur Radioastronomie,
Auf dem H\"ugel 69, 53121 Bonn, Germany }

\date{Accepted 2006 Mar 21}

\abstract{
The {\em all-sky survey of Faraday rotation}, a Key Science Project of the
planned Square Kilometre Array, will accumulate tens of millions of
rotation measure measurements toward background radio sources and will provide
a unique database for characterizing the overall magnetic geometry of
magnetic fields in galaxies and in the intergalactic medium.
Deep imaging of the polarized synchrotron
emission from a large number of nearby galaxies, combined with Faraday
rotation data, will allow us to test {\em primordial, gas flow, and
dynamo models}\ for field origin and amplification. The SKA will find the
first magnetic fields in young galaxies and determine the timescale for building up
small-scale turbulent and large-scale coherent fields. The spectrum of dynamo modes,
if existing, will be resolved. The present-day coherent field
may keep memory of the direction of the seed field which can be used for mapping
the structure of the seed field before galaxy formation.
}

\correspondence{rbeck@mpifr-bonn.mpg.de}

\maketitle
%

\section{Introduction}
\label{intro}

Galactic magnetism may have evolved in subsequent stages:\\
(1) Field seeding by primordial fields embedded in the
protogalaxy, or fields ejected into the protogalaxy by AGN jets,
radio lobes, early supernova remnants, or gamma-ray bursts.\\
(2) Field amplification by compressing or shearing flows,
turbulent flows, magneto-rotational instability, and dynamos.\\
(3) Field ordering by the large-scale dynamo.

Models referring to one or more of these stages can be tested by observations.
Radio astronomy provides the best tools to measure galactic magnetic fields.
The planned Square Kilometre Array (SKA) will allow
fundamental advances in studying the origin and evolution of magnetic
fields.

\section{Models of magnetic field origin and evolution}

\subsection{``Primordial'' models}
\label{primordial}

\begin{figure}
\centering
\includegraphics[width=0.45\textwidth]{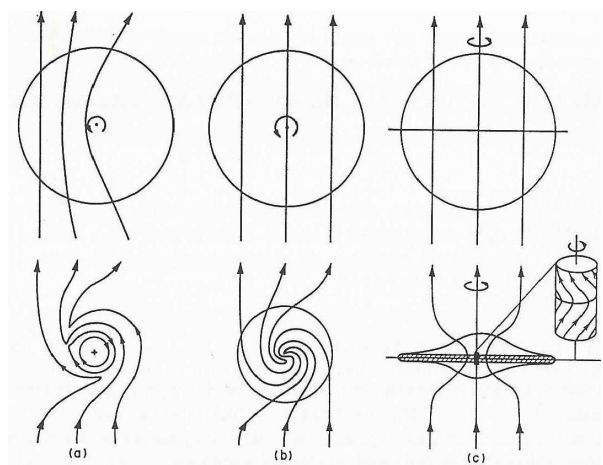}
\caption{Field structures (bottom row) generated from different
large-scale protogalactic fields (top row) in a differentially
rotating disk: (a) non-uniform field perpendicular to the rotation
axis, (b) uniform perpendicular field, (c) uniform field
parallel to the rotation axis (from Sofue \cite{sofue90}).
}
\label{sofue}
\end{figure}

A protogalactic seed field (not necessarily a primordial field
from the early Universe) was amplified
by compression during galaxy collapse and shearing by the differentially
rotating disk. To avoid winding up and field decay by reconnection,
Fujimoto \& Sawa (\cite{fuji87}) assumed field diffusion through the
disk gas. If the seed field was random, the
sheared field becomes {\em anisotropic random}. If the seed field had
a large-scale direction, the structure of the resulting field
depends on the angle between the seed field and the rotation
axis of the disk (Sofue \cite{sofue90}). A seed field perpendicular
to the rotation axis develops into a large-scale {\em bisymmetric}\
configuration in the disk (Fig.~\ref{sofue}b),
while a parallel seed field becomes {\em dipolar}\ (Fig.~\ref{sofue}c).
A non-uniform large-scale seed field may form an {\em axisymmetric}\
field in the central part of the disk (Fig.~\ref{sofue}a).

A more realistic ``primordial'' model was developed by Howard \& Kulsrud
(\cite{howard97}) who introduced coupling of the field to
gas clouds allowing for ambipolar diffusion. Any seed field
develops into an anisotropic random field with
frequent reversals on a scale of about 100~pc, the {\em coherence
length}\ of the field. Averaging the field over the galaxy generally
yields a non-zero value, a weak large-scale field.

``Primordial'' models are unrealistic because they neglect deviations from
axisymmetric rotation of the gas, such as spiral density waves.
Furthermore, such models suffer from a fundamental
problem: Shear by differential rotation increases the average field
strength, not the magnetic flux, while magnetic flux is lost into
intergalactic space by magnetic diffusion.
Estimation of turbulent diffusion leads to a decay time of the
field of only $10^8$ y (Ruzmaikin et al. \cite{ruz88}).
To maintain the field strength, field amplification by gas flows
or a dynamo is required.

\subsection{``Flow'' models}
\label{flow1}

The present-day strength and structure of galactic magnetic fields may
also be the result of {\em local}\
compression and shear by gas flows. In a kinematic
approach Otmianowska-Mazur et al. (\cite{otmia02}) showed that the
gas flow in a barred galaxy (modelled in N-body simulations) can amplify
fields and generate spiral structures with coherence lengths of a few
100~pc. Total magnetic energy and total magnetic flux decrease;
a small-scale dynamo (Sect.~\ref{flow2}) is required to maintain the field.

Dynamic MHD models of turbulent gas flows in galaxies driven by supernova
explosions were computed only for limited volumes of about 1~kpc$^3$
(Korpi et al. \cite{korpi99}; de Avillez \& Breitschwerdt\ \cite{avillez05}).
The magnetic field is sheared and compressed by the flow and reveals
a spectrum of coherent structures up to $\simeq100$~pc size.
The strongest fields are located in the regions of cold, dense gas.
This agrees well with the interpretation of the radio -- infrared
correlation (Sect.~\ref{test}).

\subsection{Dynamo models}
\label{dynamo1}

The {\em mean-field $\alpha$--$\Omega$ dynamo model}\
is based on differential rotation and the $\alpha$-effect
(Ruzmaikin et al. \cite{ruz88}; Beck et al. \cite{beck96}).
The physics of dynamo action still faces theoretical problems
(Kulsrud\ \cite{kuls99}; Brandenburg \& Subramanian \cite{brand05}).
The dynamo is the only known model which is able to generate large-scale
{\em coherent}\ magnetic fields of spiral shape.
These coherent fields can be represented as a superposition
(spectrum) of modes with different azimuthal and vertical symmetries.
In a smooth, axisymmetric gas disk the strongest mode is that
with the azimuthal mode number $m=0$ ({\em axisymmetric}\ spiral field),
followed by the weaker $m=1$ ({\em bisymmetric}\ spiral field), etc
(Elstner et al. \cite{elstner92}). These modes cause typical
variations of Faraday rotation along the azimuthal direction in a galaxy
(Krause\ \cite{krause90}). In flat, uniform disks the
axisymmetric mode with {\em even}\ vertical symmetry (quadrupole) is
excited most easily (Baryshnikova et al. \cite{bary87})
while in spherical objects the {\em odd}\ symmetry (dipole)
dominates. The timescale for building up a coherent field from a
turbulent one is $\approx 10^9$~y (Beck et al. \cite{beck94}).

Real galaxies are not uniform. Consequently, recent dynamo models include
the non-axisymmetric gas distribution in spiral arms (Moss\ \cite{moss98})
or the gas flow in a bar potential (Moss et al. \cite{moss01}),
hence combining dynamo and flow models.
Higher modes may be amplified faster than in the standard model.
Gravitational interaction with another galaxy may also modify the
mode spectrum and enhance the bisymmetric mode (Moss\ \cite{moss95}).

\section{Testing the models -- present status}
\label{test}

Total synchrotron intensity is a measure of total field strength and
density of cosmic-ray electrons (and positrons). Assuming equipartition
between the energy densities of magnetic field and cosmic rays, this
energy density is similar to that of turbulent gas motion in galaxies
(Beck\ \cite{beck05}), as imposed by dynamo models. Synchrotron intensity
is closely correlated with infrared intensity. This striking fact
tells us that cosmic-ray acceleration {\em and}\ field
amplification are continuous processes and related to star formation,
e.g. by field coupling to the ionized envelopes of cold gas clouds
(Niklas \& Beck\ \cite{niklas97}).
However, the correlation is violated for very young starburst galaxies
(Roussel et al. \cite{roussel03}). This could be due to the time needed
for the evolution of massive stars to supernova remnants, which are
believed to be the main cosmic-ray accelerators, but could also be the
result of the finite timescale for dynamo amplification.

Most galaxies reveal spiral patterns in their polarization vectors, even
flocculent or irregular galaxies. The radially decreasing pitch angles
of the observed spiral patterns agree with the predictions of
dynamo models (Beck\ \cite{beck93}; Shukurov\ \cite{shu00}).
The spiral field can be coherent or incoherent (anisotropic).
Faraday rotation measures ($RM$) are a signature of {\em coherent regular
fields}. Large-scale $RM$ patterns observed in several galaxies
(Krause\ \cite{krause90}; Beck\ \cite{beck05}) show that some fraction of the
magnetic field in galaxies has a large-scale coherent direction. The classical
case is the strongly dominating axisymmetric field in the Andromeda galaxy
M~31 (Berkhuijsen et al. \cite{berk03}; Fletcher et al. \cite{fletcher04}).
A few more cases of dominating axisymmetric fields are known (e.g. the LMC,
Gaensler et al. \cite{gaensler05}), while dominating bisymmetric fields are rare
(Krause et al.\ \cite{krause89}). The two magnetic arms in NGC~6946 (Beck \& Hoernes\
\cite{bh96}), with the field directed towards the galaxy's centre in both,
are a signature of superposed $m=0$ and $m=2$ modes. However, for most of
the (about 20) nearby galaxies for which multi-frequency observations are
available, angular resolutions and/or signal-to-noise ratios are still
insufficient to reveal a mixture of magnetic modes -- if existing.

\begin{figure}
\centering
\includegraphics[bb = 62 135 543 645,width=0.42\textwidth,clip=]{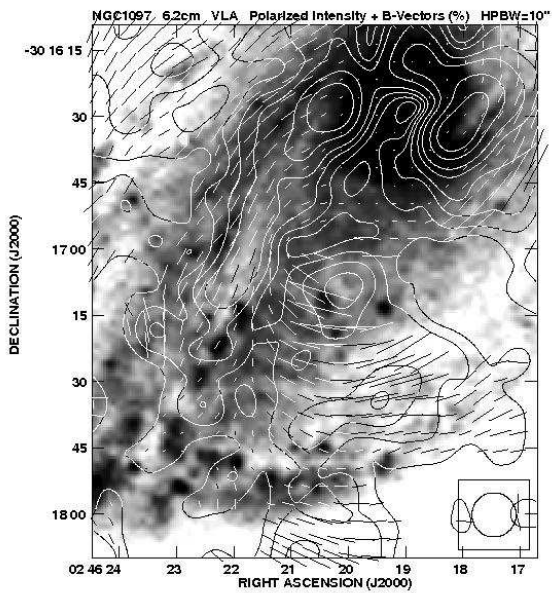}
\caption{Polarized intensity contours and observed
   $B$-vectors ($E+90^\circ$) of the central and southern parts of
   NGC~1097 at $\lambda6.2$~cm at 10\arcsec\ resolution.
   The background optical image was kindly provided by Halton Arp
   (from Beck et al. \cite{beck+05}).}
\label{n1097}
\end{figure}

Polarization angles are ambiguous by $\pm 180^{\circ}$ and hence
insensitive to field reversals. Compression or stretching of turbulent
fields with random orientations generates incoherent {\em anisotropic random}\
fields which reverse their direction frequently within the telescope beam, so
that Faraday rotation of the extended polarized emission
is small while the degree of polarization can still be high.
Observations of several spiral galaxies revealed a highly ordered spiral
pattern in the polarization (B-) vectors, but small Faraday rotation,
so that anisotropic random fields seem to dominate over the coherent
regular ones. Such fields are predicted by flow models (Sect.~\ref{flow1}).
A striking example is the barred galaxy NGC~1097
where the high degree of polarization along the southern bar is mostly due
to compressed random fields (Fig.~\ref{n1097}). Ram pressure from the
intracluster gas or interaction between galaxies have a similar compressional
effect (Chy\.{z}y et al., this volume; Chy\.{z}y \& Beck\ \cite{chyzy04}).
Dwarf irregular galaxies with almost chaotic rotation host
turbulent fields with strengths comparable to spiral galaxies, but no
large-scale coherent fields (Chy\.{z}y et al. \cite{chyzy03}).

The structure of coherent (regular) fields may extend to small scales
which cannot be resolved by present-day observations, e.g.
elongated field loops in the disk or Parker loops.
Observations at higher resolution are needed to distinguish between
anisotropic random and unresolved regular fields.

Present-day observations are limited by sensitivity at high resolution.
Polarized intensities are low and can be detected only with large
single-dish telescopes (Effelsberg, GBT) and with the VLA and ATCA in
compact configurations. Hence, the best available spatial resolutions for
polarization and Faraday rotation studies are only 300--500~pc
in the nearest spiral galaxies. Galaxies beyond $\simeq20$~Mpc
distance cannot be sufficiently resolved in polarization.

$RMs$ toward polarized, point-like background sources do not suffer
from limited resolution. Such data are scarce due to limited
sensitivity. As their number increases with the angular size of the
galaxy, only the largest ones were studied so far.
Han et al. (\cite{han98}) found 21 polarized sources behind M~31,
Gaensler et al. (\cite{gaensler05}) about 100 sources behind the LMC.

\section{Testing the models with the SKA}

The {\em all-sky survey of Faraday rotation}, a SKA Key Science Project,
will accumulate tens of millions of rotation measure ($RM$) measurements
toward background radio sources (Gaensler et al. \cite{gaensler04}).
This will provide a unique database for understanding the structure
and evolution of magnetic fields in galaxies and in the intergalactic
medium (Beck \& Gaensler\ \cite{beck+04}). {\em Faraday tomography}\
of the Milky Way at $\le1$~GHz will yield a three-dimensional picture
of the magnetic field within a few kpc of the Sun. High-resolution
synchrotron imaging at $\ge5$~GHz of a large number of nearby galaxies,
combined with $RM$ data, will allow us to determine their magnetic field
structure, and to test both the dynamo and primordial field theories
for field origin and amplification.

Typical polarization intensities of nearby galaxies at 5~GHz are
$\sim$0.1~mJy per $15''$ beam. Within a $1''$ beam, $\sim0.4~\mu$Jy is
expected which the SKA can detect in $\sim1$ hour of integration.
This will allow polarization and Faraday rotation mapping in galaxies
out to a distance of about 100~Mpc. Furthermore, such observations
will reveal a large number of $RMs$ toward background sources which
can be used for an independent investigation of the detailed field
structure.

\subsection{Primordial against dynamo models}
\label{dynamo2}

A large sample of data on the total and polarized synchrotron intensity
and Faraday rotation in young galaxies will clarify the timescales for the
generation of large-scale coherent and of small-scale
fields, to be compared with the models. The SKA will
also provide $RM$ data toward sources behind galaxies with little or no star
formation, like dwarfs and ellipticals, where no synchrotron emission is
detectable but magnetic fields may exist, triggered by turbulence driven
by type I supernovae (Moss \& Shukurov\ \cite{moss96}).

The SKA will confidently determine the Fourier spectrum of
dynamo modes. The azimuthal mode of order $m$ has $2m$ reversals and
can be detected if $\simeq10(m+1)$ independent azimuthal sectors
are resolved in the disk of the galaxy. Let $\Theta$ be
the telescope's angular resolution, $R$ the mean radius of polarized
emission, $i$ the disk's inclination ($i=0$ for face-on) and
$D$ its distance. The highest resolvable mode is
$m_{max} \approx \pi R \cos{i}/5\Theta D-1$. To resolve all modes up to
$m=4$ in a galaxy of 5~kpc radius and $45^o$ inclination at a distance of
100~Mpc, a resolution of $\approx 1''$  at $\ge5$~GHz is
required which the SKA will provide with high signal-to-noise ratio.
As the average $RM$ signal from a coherent field
parallel to the disk varies with $\sin{i}$, mildly inclined galaxies
are preferable.

The SKA has the potential to increase the galaxy sample with well-known
field patterns by up to three orders of magnitude. The conditions for
the excitation of dynamo modes can be clarified. For example, interactions with
companion galaxies may enhance the bisymmetric $m=1$ mode (Sect.~\ref{dynamo1}).
A dominance of bisymmetric fields for non-interacting
galaxies would be in conflict with existing dynamo models and would
support the primordial field origin (Sect.~\ref{primordial}).

Galactic dynamo models also predict the preferred generation of quadrupolar
patterns (Sect.~\ref{dynamo1}) where the field in the disk has the same
sign above and below the plane. Primordial models predict dipolar patterns
with a reversal in the plane (Fig.~\ref{sofue}c)
which can be distinguished by observing $RMs$ in edge-on galaxies.
However, their polarized emission is weak in the disk (due to strong
depolarization) and also in the halo (strong energy losses of the cosmic-ray
electrons). The determination of the global vertical field symmetry has not
been possible yet. This experiment also must await the SKA.

$RMs$ toward background sources will allow us to trace coherent fields to
large galactic radii and hence to derive restrictions for dynamo action.
If the $\alpha$ effect is driven by supernova remnants or by Parker loops
(Hanasz, this volume), dynamo modes should be excited
preferably in the star-forming regions. If the magneto-rotational
instability is the source of turbulence and of the $\alpha$
effect (Sellwood \& Balbus\ \cite{sell99}; Kitchatinov \& R\"udiger\
\cite{kit04}), field amplification will be seen out to large galactic radii.

\subsection{Large-scale seed fields}
\label{seed}

\begin{figure}
\centering
\includegraphics[width=0.38\textwidth]{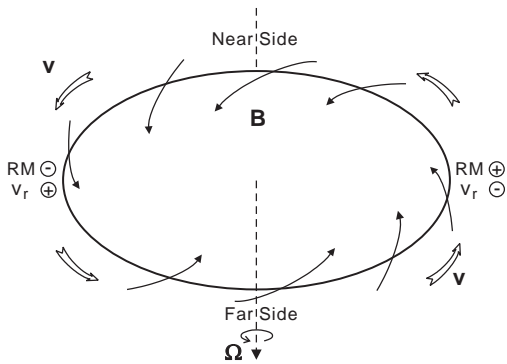}
\caption{Inward-directed axisymmetric magnetic field in an
inclined galaxy with trailing spiral arms. The signs of Faraday rotation
measure $RM$ and rotational velocity $v_r$ along the line of sight are
indicated at the major axis (from Krause \& Beck\ \cite{krause98}).}
\label{rm}
\end{figure}

Most of the (few) galaxies known to host a dominating axisymmetric $m=0$
mode possess a radial field component, which is directed {\em inwards}\
(Krause \& Beck\ \cite{krause98}).
Dynamo models preserve the memory of the direction of
the large-scale seed field. The sign of the radial
field component follows from the observed Faraday rotation
($RM$) and rotational velocity along the line of sight ($v_r$) on
the major axis (Fig.~\ref{rm}): opposite signs
of $RM$ and $v_r$ indicate an inward-directed field, same signs an
outward-directed field. SKA's sensitivity and broad frequency bands
will allow to observe galaxies out to more than 100~Mpc distance
and to map the structure of the large-scale seed field before
galaxy formation.

\subsection{Gas flow and small-scale dynamo}
\label{flow2}

The failure to detect a coherent magnetic field in a resolved galaxy
and the detection of predominantly turbulent fields with the SKA
would indicate that mean-field dynamo action (Sect.~\ref{dynamo1}) is
unimportant (e.g. due to its long timescale) and that gas flows structure
the field, supported by the {\em small-scale}\ or {\em fluctuation dynamo}
(Subramanian\ \cite{subra98}; Brandenburg \& Subramanian\ \cite{brand05},
see also this volume), which amplifies turbulent, incoherent magnetic fields,
does not rely on differential rotation, and can work in all galaxy types.

\end{document}